\documentstyle[12pt,aasms4]{article}

\begin{document}

\title{Inclination Effects in Spiral Galaxy Gravitational Lensing}

\author{Ariyeh H. Maller\altaffilmark{1}, Ricardo A. Flores\altaffilmark{2}
 and Joel R. Primack\altaffilmark{1}}
\altaffiltext{1}{Physics Department, University of California, Santa Cruz,
 CA 95064}
\altaffiltext{2}{Physics Department, University of Missouri, Saint Louis,
 MO 63121}

\begin{abstract}

Spheroidal components of
spiral galaxies have been considered the only
dynamically important component in gravitational lensing studies thus far.
Here we point out that including the disk component can have a significant
effect, depending on the disk inclination, on a variety of lensing properties
that are relevant to present studies and future surveys. As an example, we
look at the multiple image system B1600+434, recently identified as being
lensed by a spiral galaxy. We find that including the disk component one can
understand the fairly large image separation as being due to the inclination
of a typical spiral, rather than the presence of a very massive halo. The
fairly low magnification ratio can also be readily understood if the disk is
included.
We also discuss how such lensed systems might allow one to constrain
parameters of spiral galaxies such as a disk--to--halo mass ratio, and disk
mass scale length. Another example we consider is the quasar multiple--lensing
cross section, which we find can increase many--fold at high inclination for
a typical spiral. Finally, we discuss the changes in the gravitational lensing
effects on damped Lyman alpha systems (DLAS) when disk lensing is included.

\end{abstract}

\keywords{gravitational lensing---galaxies:structure---galaxies:spiral
---quasars:absorption lines}

\section{Introduction}
Galaxies responsible for gravitational lensing have typically been modeled as
one-component
systems in which the entire mass distribution of the galaxy lens, visible plus
dark matter, is fairly well approximated by a singular isothermal sphere (SIS):
$\rho(r)=\rho_o (r_o/r)^2$ (see e.g. Narayan and Bartelmann
\markcite{NB}1997).
This is probably justified for E/SO galaxies where
both components have a surface density fairly independent of orientation (a
small ellipticity is required only to understand the presence of four-image
systems) and the high central density represents the high density of the
visible matter. By contrast, in spiral galaxies the disk component has a very
different structure from the dark matter halo and the bulge, which results
in significantly different projected surface density depending on orientation.
Given that the projected surface density of the disk changes by a factor of
radius/thickness $\sim 15$  between a face-on and an edge-on spiral,
and that the dark and visible components are known to contribute roughly
equally to the total mass within 4 -- 5 disk scale lengths, one would
expect significant effects on lensing properties for images within
a couple of optical radii.

The identification of the lensing galaxy in the double image
system B1600+434 as a nearly edge on spiral by Jaunsen and Hjorth
\markcite{JH}(1997)
gives us a perfect example to demonstrate the importance of including the disk
component in modeling lensing by spirals.
However, an understanding of the lensing geometry of spirals is important for
several reasons besides that of modeling observed systems.  First,
recently the gravitational lensing effects on damped Lyman alpha systems
(DLAS) have been investigated (Smette, Claeskens, and Surdej\markcite{SCS}
1997; Bartelmann and Loeb \markcite{BL}1996).
At least some DLAS are believed to be due to absorption in protogalactic
disks or spiral
galaxies (Wolfe \markcite{W}1988, \markcite{W2}1997), and if the gas 
column density falls off
exponentially, one is necessarily dealing with lensing within a few exponential
scale lengths of the galaxy center, which is where we expect disk lensing to
be important. Secondly, the extra parameters added by introducing a disk
component are the physically interesting ones of disk surface density and
scale length,  so that the observational study of gravitational lensing by 
spirals will provide
new constraints on the distribution of mass in the various components of
spirals.  Although only a few such systems are currently known, surveys such
as the Sloan Digital Sky Survey will catalog a large number of quasars, among
which there should be a significant number lensed by spiral galaxies.
This is why it is also important to understand the effect of disk lensing
on statistical properties such as the cross section for multiple-image
lensing by spirals.

In this paper we first explain our method of modeling spiral galaxy lenses as
singular isothermal spheres plus infinitesimally thin disks (SIS+disk).
We start with the simple example of a constant density disk, to show the
effect of adding a disk.
We show that such a model fits many of the observations on B1600+434,
and discuss the results of modeling, more properly, with an exponential disk.
We then discuss the dependence of the lens cross section for multiple lensing
of background sources on the inclination of the disk, and discuss also how it
would enhance the gravitational lensing effects on DLAS.
We finish with the conclusions and implications from our work.

\section{The Model}

We model a spiral galaxy as consisting of two mass components: a spherical 
halo and an infinitesimally thin disk.  We assume that the halo has the 
density profile of a singular isothermal sphere, while the disk
is treated as having surface density that is either constant, or falls
off exponentially. For simplicity and clarity we start by treating the
case of the constant density disk.

The orientation of the disk defines the coordinate system, such that
the x and y directions are along the major and minor axis,
respectively, of the projected disk, and the disk center defines the
origin.  The model has then 6 parameters:
the position of the source ($\beta_x,\beta_y$), the inclination of the
disk $\gamma_o$, the velocity dispersion of the SIS $\sigma_v$ (the SIS can 
be equivalently be described by $\rho(r)=\sigma_v^2/2G\pi r^2$), and
the central surface density $\Sigma_o$ and characteristic length of
the disk.  For a constant density disk this characteristic length is
the radius of the disk, while in the exponential disk case this
becomes the scale length, $R_s$.  Thus, measuring the position of the
lens, the inclination of the disk, the two image positions
($\theta_x,\theta_y$), and the magnification ratio completely
determines the system.

The inclined circular disk projects onto an ellipse.  We denote the axis
ratio of an ellipse in general as  $\gamma= b/a$, and that of our
disk as $\gamma_o$.
Any point $(\theta_x,\theta_y)$ external to the ellipse will define a new
confocal ellipse with axis ratio $\gamma'$. Then the reduced
deflection angle, modified from Schramm \markcite{Sc}(1990), is given by
\begin{eqnarray}
        \alpha_x =\frac {2\Sigma_o}{\Sigma_{cr}} \theta_x \frac{1-\gamma'}
{1-\gamma_o^2}
\\
        \alpha_y =\frac {2\Sigma_o}{\Sigma_{cr}} \theta_y \frac{1-\gamma'}
{\gamma'(1-\gamma_o^2)} .
\end{eqnarray}
Here $\Sigma_{cr}$ is the usual critical density defined as
\begin{equation}
 \Sigma^{-1}_{cr}=\frac {4\pi G D_l D_{ls}}{c^2D_s} 
\end{equation}
and $D_i$ is the lens, source, or lens-to-source angular diameter 
distance.
If the point $(\theta_x,\theta_y)$ is inside the disk, then $\gamma'$ is just
replaced with $\gamma_o$.

\section{The System}

In the gravitational lens system B1600+434 identified by Jackson
et al. \markcite{J}(1995), the lensed quasar is at a redshift $z_s=1.61$, the
image separation is 1\farcs4, and the intensity ratio is
${I_A}/{I_B}=1.3$ in the radio (in the optical, image B suffers significant 
extinction).  The lens galaxy, identified as a nearly edge on spiral by Jaunsen
and Hjorth \markcite{JH}(1997), has an estimated photometric redshift $z_l \sim
0.4$, and axis ratio $a/b = 2.4 \pm 0.2$, or $\gamma_o=0.42$. (It is possible 
that this galaxy is actually an S0, as suggested to us by C. Kochanek,
private communication, Jannuary 1997, but this will not effect our analysis
as there is still a disk.)
To emphasize that this paper is meant as a qualitative exploration of the
SIS+disk model, and because $z_l$ is still uncertain, we treat
all measurements as exact and include no estimates of errors.  The
positions of the images in the lens-centered, axes-defined coordinate
system are given in Table I.
Note that the lensing galaxy is just 0\farcs35 from image B.  In all of our 
calculations, we assume $\Omega=1$, $q_o=0.5$ and $H_o=100h\,  {\rm kms}^{-1} 
{\rm Mpc}^{-1}$.  Also we calculate angular diameter distances assuming
all the matter in the universe is smoothly distributed 
(see Schneider, Ehlers and Falco \markcite{SEF}1992).

Treating the lens as a singular isothermal sphere
requires a velocity dispersion of 200 km/s to account for the image separation.
Of course, in this case the lens must be collinear with the images, which is
not the case observationally (one would need to introduce some ellipticity in
the dark matter distribution to avoid this). However, we can obtain an estimate
of the implied relative magnification and the time delay by shifting the lens
to the collinear position along a line perpendicular to $\overline{AB}$. 
The implied
time delay is $\Delta t_{AB}=24h^{-1}$ days, and the magnification ratio is
3.8. (The results are similar if the lens position is shifted keeping the
ratio of
image-lens distances: $\Delta t_{AB}=22h^{-1}$, magnification ratio 3.2.)
If one chooses instead to shift the lens to a position that gives the
observed relative magnification, the time delay is $\Delta t_{AB}=5.3h^{-1}$
days, but then the images would both be at a similar distance from the lens, 
which is not as observed.

If instead we model the system as a SIS+constant density disk, we introduce
two new parameters, surface density and disk length. The farther image's
light passes through the disk only $8.7h^{-1}$ kpc from its center, so we can
start by assuming that the disk length is greater than this and remove one
parameter.  The equations for a disk with radius greater than $8.7h^{-1}$
become
\begin{eqnarray}
\alpha_x =\frac {2\Sigma_o}{\Sigma_{cr}} \frac{\theta_x}
{1+\gamma_o}
\\
\alpha_y =\frac {2\Sigma_o}{\Sigma_{cr}} \frac{\theta_y}
{\gamma_o(1+\gamma_o)}
\end{eqnarray}
and the lensing potential of the disk in this case is
\begin{eqnarray}
\Psi = \frac {\Sigma_o}{\Sigma_{cr} (1+\gamma_o)} (\theta_x^2 +
\frac {\theta_y^2}{\gamma_o}) .
\end{eqnarray}
Solving for the image positions in Table I gives $\Sigma_0=7.8\times10^8
h{M_{\sun}}/{\rm kpc^2}$, $\sigma_v=135$ km/s, and a source position of
(0.037, 0.12).  The magnification ratio is 1.3, and the time delay is now
$\Delta t_{AB}=68h^{-1}$ days. See Figure 1. 

The addition of the disk component has four main effects.  
First it breaks the 
spherical symmetry, allowing the lens not to be collinear with the images.
Second, the velocity dispersion in the halo becomes significantly less,
as the disk component contributes substantially to the deflection angle.
And third, in this case the magnification ratio is reduced, though in 
general it depends strongly on the orientation of the images around 
the lens axis.
Also the time delay is increased, because the potential now has a term 
quadratic in $\vec{\theta}$, and because the geometric time delays
do not cancel out as they do in the SIS case.

It is interesting to note that the surface density of the disk comes
out as a reasonable number, although a priori it could have been
anything.  The deflection caused by a disk with density $\Sigma_o \sim
10^8 h^{-1}{M_{\sun}}/{\rm kpc^2}$ is of the same scale as that caused by
a halo with $\sigma_v \sim 100$ km/s.  Since we observe roughly this
amount of luminous matter in the disk, it is clear that the disk's
contribution cannot be ignored.

Of course a disk with constant surface density out to 9$h^{-1}$ kpc is not a 
realistic model.  We can realistically assume the disk's surface density 
falls off as an exponential with scale length $R_s$.  In Figure 2 the images
from an exponential disk with $R_s=5h^{-1}$ kpc,
$\Sigma_o=1.9 \times 10^9 h{M_{\sun}}/{\rm kpc^2}$ and a halo with
$\sigma_v=120$ km/s are shown.  The images are within the position errors
for the lens center, and the magnification ratio is roughly correct at 0.97.
	
Interestingly we find that the magnification ratio is strongly dependent 
on the relative orientation of the images and the lens axis.  Figure 3 shows
the magnification ratio as a function of angle from the y-axis.
A magnification ratio of 1.3 requires that the source be about $20\arcdeg$
off the y-axis, which also corresponds to the image orientation found in
B1600+434. 

In this preliminary study we
have found a surprisingly large disk-to-halo ratio in order
to fit the image positions. For the exponential disk, the total
disk mass is $3\times 10^{11} h^{-1}M_{\sun}$, compared to a halo 
velocity dispersion of $\sigma_v=120$ km/s. 
In comparison, standard lore (Binney and Tremaine\markcite{BT} 1987)
gives values of 
$6\times 10^{10} M_{\sun}$ for the disk mass and 155 km/s velocity dispersion
for the Milky Way with a disk scale length of 3.5 kpc. While there are
large uncertainties in these parameters for the Milky Way, our calculated
disk mass does seem excessively large.

Without an understanding of the uncertainties, it is 
hard to judge how significant these results are. 
In particular the inclination we have used of $65\arcdeg$ (based on the axis 
ratio measured by Jaunsen and Hjorth\markcite{JH} 1997) could be much
higher, which tends to decrease the scale length and hence the total disk
mass.
A preliminary inspection of HST archival images of B1600+434 reveals a dust
lane nearly bifurcating the disk, suggesting an inclination greater than 
$80\arcdeg$.  However, if the large disk mass did 
turn out to be significant, it may be evidence that the dark matter halo is 
flattened (see for example Olling\markcite{O} 1996; Sackett\markcite{Sa} 1996; 
Rix\markcite{R} 1995),
and the non-spherically situated dark matter is  
contributing to what we have assumed is a visible matter disk.  It is also
possible that the dark matter halo may have a core radius, and since the 
lensing is mostly sensitive to the matter inside the images, the 
$\sigma_v$ we are measuring is lower than the value one would use to 
characterize the galaxy. We are preparing to address these issues in a more
detailed study to follow.

\section{Other effects of Including a Disk Component}
The inclination of the disk can have a dramatic effect on the lens cross
section for multiple lensing of background sources, as we show in Figure 4.
Here we model the disk with constant surface density again, but we choose
parameters more representative of an $L_{*}$-type spiral.

In this case it is possible to obtain a simple expression for the critical
inclination, $\gamma_c={\rm cos}\theta_c$, at which the cross section begins 
to exceed that of
the SIS halo alone. Just note that this occurs when the inner caustic due to
the disk touches the SIS caustic. If the inclination slightly exceeds
$\theta_c$, then we can have a source just outside the SIS caustic and just
inside the disk caustic (see the $75\arcdeg$--curve in Figure 4). A source at
this location, $(\beta_x,\beta_y)=(\alpha_{SIS},0)$ with
$\alpha_{SIS}=(D_{ls}/D_{s})4\pi(\sigma_v/c)^2$, has three images.
Using the lens equation
\begin{eqnarray}
\vec{\beta}=\vec{\theta}-\vec{\alpha}_{disk}-\vec{\alpha}_{SIS}
\end{eqnarray}
we see that, assuming the disk radius is larger than $\alpha_{SIS}$,
 one image is at
\begin{eqnarray}
\theta_{xo} = 2\alpha_{SIS}\Bigg/ \left(1-\frac{2\Sigma_o}{\Sigma_{cr}
(1+\gamma_c)} \right)\ {\rm and}\ \theta_y = 0.
\end{eqnarray}
Because we know that as the source passes through the caustic the images 
will merge, we know that the other two images must be close to the first one.
Thus there are two images at $\theta_x \approx \theta_{xo}$ and  
$\pm \theta_{yo}$, where $\theta_{xo} \gg \theta_{yo}$. 
For these two images the lens equation requires
\begin{eqnarray}
{\alpha_{SIS}\over{\theta_{xo}}}+
{2 \Sigma_o\over\Sigma_{cr}\gamma_c(1+\gamma_c)} = 1
\end{eqnarray}
independently of $\theta_{yo}$.
Therefore,
\begin{eqnarray}
\gamma_c= {\rm cos}\theta_c = {1\over{2}}\left(\sqrt{
1+20\frac{\Sigma_o}{\Sigma_{cr}}+\left(\frac{2\Sigma_o}{\Sigma_{cr}}\right)^2}
-1-2\frac{\Sigma_o}{\Sigma_{cr}}\right)
\end{eqnarray}
gives the critical inclination.
  
In Figure 4 we show two
inclinations for a disk of mass $M_{disk} = 10^{11} h^{-1}M_{\sun}$ and radius
$R_{disk} = 8 h^{-1}$ kpc, inside a SIS halo with $\sigma_v=155$ km/s.
For this choice of parameters $\theta_c \approx 73\arcdeg$. Thus, at an
inclination of $75\arcdeg$ the cross section shows a modest increase.
By contrast, at $85\arcdeg$ we find a very large increase by a factor of
$\approx 3.5$ over that of the SIS halo alone.
The disk model is not realistic, of course, but
the effect is quantitatively similar for an exponential disk. In particular,
$\theta_c$ is still fairly well approximated by Eq. (10). 

Whether this increase in cross section significantly increases the 
fraction of spiral galaxies responsible for multiple image lensing depends on
the amount of mass in the disk.  In Figure 5 one sees that for parameters like
those quoted for the Milky Way the average (integrated over cos$\theta_i$)
increase is small,
roughly $\sim 25\%$. While, for a heavy disk like our model of B1600+434, 
the average cross section increases by $\sim 90\%$.  But since the 
amplification also increases with disk 
inclination, the brightest quasars lensed by disk galaxies will probably have
highly inclined disks, especially if the disks are massive. 
Thus the fraction of lenses that are spirals,
and their relative inclinations, will place strong constraints on the total
disk mass.
We shall discuss these issue in more detail in a subsequent
publication.

The last problem we study here is how our results would affect the
gravitational lensing effects on DLAS. Bartelmann and Loeb\markcite{BL}
(1996) and Smette \markcite{Sm}(1995) have pointed out that gravitational
lensing could significantly affect the observed properties
of DLAS for two reasons. First, magnification bias significantly
enhances the probability of observing damped absorption in a QSO spectrum.
Second, there is a cutoff in the observed column density due to light bending,
which makes it far more likely to observe absorption that occurs at large
radii. Since the column density of absorbing HI falls off exponentially, this
significantly affects the observed column densities.

In their studies, Smette et al. \markcite{SCS}(1997), 
Bartelmann and Loeb \markcite{BL}(1996) treated the spiral
disk as
essentially massless for lensing purposes. We find that including the disk
lensing will significantly enhance both these effects, as we show in Figure 6.
There we show the effects of adding disk lensing on the magnification and
impact parameters of two sources. First, note that there is a significant
increase in the magnification of the image that, precisely, occurs on the
surface of the disk. For the example shown, the $L_*$--type disk considered
above, the point-source magnification increase is $\approx 5.2 (3.0)$ 
for the source at
position 1 (2). Second, there is an increase in the impact parameter, which
would result in further reduction of the observed column density. These
results are typical of what we find at high inclinations of $60\arcdeg$--
$80\arcdeg$.  Note that the increased magnification by inclined disks
increases the
likelihood of lensing in this configuration; this could obviate the
need for rather thick disks (e.g., Wolfe\markcite{W2} 1997) to account
for the metal line systems seen in DLAS.
Also one must be aware that unresolved multiple images may pass through 
the same disk at various places complicating the reconstruction of the 
absorbing system.

A complete analysis is now far more involved, as one must model the disk as an
exponential disk, average over the observed magnification, and average over
the disk inclination. 
Also the effects of dust in obscuring images (as clearly demonstrated in the
relative reddening of image B to A in B1600+434) must be included.
We will discuss the full problem in a subsequent publication.

\section{Conclusions}

We have shown that the inclusion of a disk component in a spiral gravitational
lens model 
has significant effects that cannot be ignored. Our results imply that a host
of issues ranging from the ability of spirals to lens QSO's to detailed
properties of lens systems should be reconsidered to properly include the
lensing effects of their disks. We have shown, as an example, that a SIS+disk
model can account for the observations of the B1600+434 system with fairly
reasonable parameters. The implications of our results are that a velocity
dispersion $\sigma_v$ significantly smaller than the SIS prediction of
200 km/s, and a time delay greater than one month, are to be expected.

With more data and more detailed analysis, the modeling of multiply--imaged
systems lensed by spiral galaxies will constrain the relative amounts of 
matter in the disk and halo when disk lensing is included. However, because
of the dependence on new parameters, further study will be necessary before
such systems can give an unambiguous measurement of the Hubble parameter $h$.
We have also noted that the disk contribution to lensing will make the
amplification and by-pass effects on the observed distribution of DLAS in
bright QSO's even more pronounced at high column densities.

After the first version of this paper was posted on the preprint archive
as astro-ph/9701110, the preprint astro-ph/9702078 was posted by Wang 
and Turner discussing critical curves and caustics of inclined,
constant density disks.  Their equations and figures can be seen
to be the same as ours if one is careful to scale appropriately by the Hubble
constant $h$, and to notice that they have calculated angular diameter 
distances with all the matter in the universe in clumps while we have 
assumed it to be distributed smoothly.

This work has been supported by NSF grant PHY-9600239 at
UM-St. Louis, and by NSF grant PHY-9402455 and a NASA ATP grant at
UCSC. AM also acknowledges GAANN support.  The authors are grateful for 
helpful suggestions from the referee, Matthias Bartelmann; for help from 
Raja Guhathakurta and Steve Vogt in interpreting the HST archival image of
B1600+434; and correspondence from A.O. Jaunsen, Chris Kochanek, and Tom 
Schramm.

\newpage

\figcaption[fig1.ps]{Lensing model for B1600+434 by a SIS + {\bf uniform disk},
with disk length $>8.7 h^{-1}$ kpc. The observed images are the black dots. 
The small contours show the images of a small circular source 
located at the star symbol in the source plane.  
Since lensing conserves surface brightness the relative sizes of the images
is indicative of their magnification ratio.  The model magnification ratio
calculated from point sources is 1.3.  The source position is chosen such 
that the model images are centered on the observed images. The model 
magnification ratio is roughly equal to 1.3.  Here $\sigma_v=135$ km/s and
$\Sigma_o=7.8 \times 10^8 h {M_{\sun}}/{\rm kpc^2}$. The dashed
curve is the SIS ``caustic''. The model lens has two caustics: a ``tangential''
caustic (dotted line), and a ``radial'' caustic that coincides with the
SIS caustic. Also the critical curve corresponding to the ``tangential''
caustic is shown as the outer solid curve.}

\figcaption[fig2.ps]{Lensing model for B1600+434 by SIS + {\bf exponential
disk}.   The observed images are the black dots. The images are only slightly
off their observed positions and have roughly the correct magnification 
ratio 0.97. $R_s= 5 h^{-1}$kpc, 
$\Sigma_o = 1.9\times 10^9 h {M_{\sun}}/{\rm kpc^2}$
and $\sigma_v=135$ km/s. Caustics are as in Figure 1.}

\figcaption[fig3.ps]{Dependence of magnification on source orientation. The 
magnification ratio is plotted vs. angle from the y-axis.  The source is
at a constant distance of $0\farcs126$.  The lens
parameters are the same as in Figure 1.}

\figcaption[fig4.ps]{Dependence of multiple-lensing cross section on disk
inclination.
Here the lens has a halo with $\sigma_v=155 h$ km/s and a uniform
density disk with mass $M_{disk} = 10^{11} h^{-1}M_{\sun}$ and radius
$R_{disk} = 8 h^{-1}$ kpc. The dashed circle represents the caustic of the 
SIS halo.  When inclined at $75\arcdeg$ the ``tangential'' caustic is the 
thick line 
and the ``radial'' caustic coincides with the dashed circle.  Inclined at 
$85\arcdeg$ the ``tangential'' caustic is the thin line and the ``radial''
caustic is the dotted line that joins the dashed circle.
(A source behind the tip of the arrow-like region would have 6 images.)}

\figcaption[fig5.ps]{Increase in multiple image cross section to SIS alone 
as a function of inclination for a SIS + exponential disk model.  
The solid line is for a massive disk like that found for B1600+434, see Figure
2.  Broken line is for a less massive disk like a Milky Way type galaxy
with $M_{disk} = 6\times 10^{10} h^{-1}M_{\sun}$ and radius $R_{disk} 
= 3.5 h^{-1}$ kpc.}

\figcaption[fig6.ps]{Dependence of magnification and bending angle on disk
inclination. The solid lines are image contours of two circular sources
(1 and 2) at positions marked by the star symbols. Here the disk is
as in Figure 4, inclined at $75\arcdeg$. The dashed contours mark where the
image contours lie for the SIS halo alone. The dotted line marks the edge of
the constant density disk.  See discussion in the text.}
\newpage
 
\begin{deluxetable}{cll}
\footnotesize
\tablecaption{Image and Lens Positions \label{tbl-1}}
\tablewidth{0pt}
\tablehead{
\colhead{Object} & \colhead{$\theta_x$}   & \colhead{$\theta_y$}}
\center
\startdata
Lens 	&	0\farcs0 	&	0\farcs0 	\nl
A	& 	-0\farcs08	&	1\farcs12 	\nl
B	&	0\farcs26	&	-0\farcs23	\nl
\enddata
\end{deluxetable}

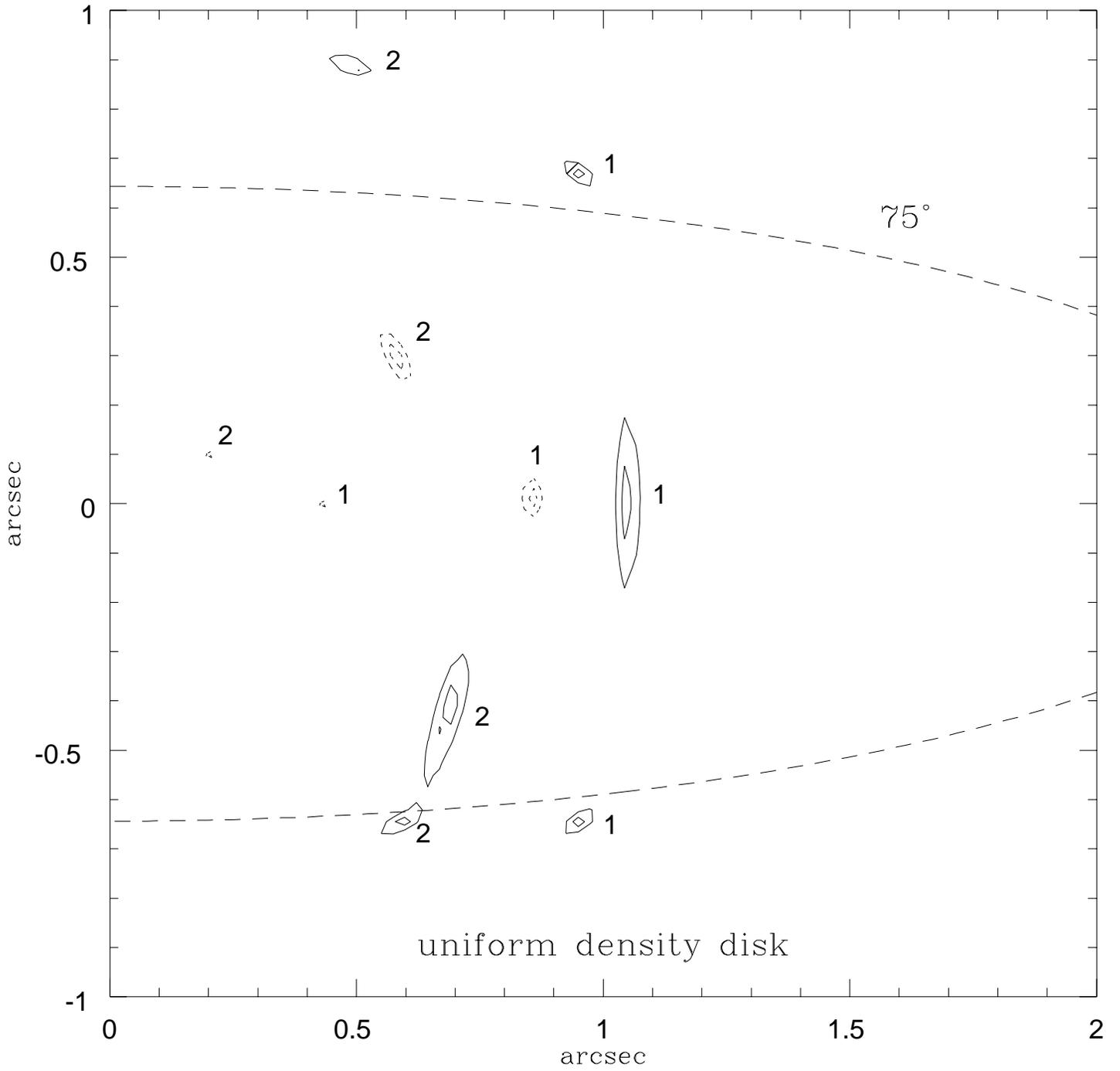
\end{document}